\documentstyle[psfig,amssymb]{mn}
\title[The Phoenix Deep Survey: the radio properties of the hard X-ray selected sample]
{The Phoenix Deep Survey: the radio properties of the hard X-ray selected sample}

\author[A. Georgakakis et al.] {A. Georgakakis$^{1}$\thanks{email: age@astro.noa.gr},
  A. M. Hopkins$^{2}$\thanks{Hubble Fellow},  
  J. Afonso$^3$, M. Sullivan$^{4,5}$, B. Mobasher$^6$,  \\ \\
  {\LARGE L. E. Cram$^{7,8}$} \\ \\
  $^1$ Institute of Astronomy \& Astrophysics, National Observatory of
  Athens, I. Metaxa \& B. Pavlou, Penteli, 15236, Athens, Greece \\
  $^2$  Department of Physics and Astronomy, University of Pittsburgh,
  3941 O'Hara Street, Pittsburgh, PA 15260, USA \\
  $^3$ Centro de Astronomia e Astrof\'{\i}sica da Universidade de Lisboa,
   Observat\'orio Astron\'omico de Lisboa, Lisboa 1349-018, Portugal\\  
  $^4$ Department of Astronomy and Astrophysics, University of Toronto, 60 St. George Street, Toronto, ON M5S 3H8, Canada\\
  $^5$ Physics Department, University of Durham,  Science Labs,
  South Road, Durham, DH1 3LE \\
  $^6$ Space Telescope Science Institute, 3700 San Martin Drive,
 Baltimore, MD 21218, USA \\
  $^7$ Australian Research Council, GPO Box 9880, Canberra ACT 2601,
  Australia \\
  $^8$ The Australian National University,  Canberra ACT 0200,
  Australia\\ 
}

\begin{document}
\maketitle  

\begin{abstract}
  The radio properties of hard (2-8\,keV) X-ray
  selected sources are explored by combining a single 50\,ks XMM-{\it
  Newton} pointing with the ultra-deep and homogeneous Phoenix radio 
  (1.4\,GHz) survey (Hopkins et al. 2003).  A total of 43 sources are
  detected above the X-ray flux limit $f_X(\rm 2 - 8 \, keV) = 7.7
  \times 10^{-15}\, erg\, s^{-1}\, cm^{-2}$, with 14 of them
  exhibiting radio emission above $\rm \approx40\,\mu Jy$ ($3\sigma$).
  The X-ray/radio matched population lies in the borderline between
  radio loud and  quiet AGNs and comprises sources with both soft
  and hard X-ray spectral properties suggesting both obscured and
  unobscured systems. The spectroscopically identified sub-sample (total
  of 6 X-ray/radio matches) comprises narrow emission line AGNs (4)
  with hard X-ray spectral properties and broad line sources (2) with
  soft X-ray spectra. We find evidence that the fraction of
  X-ray/radio matches increases from  $\approx20\%$ for sources with
  rest-frame column density $\rm N_H<10^{22}\,cm^{-2}$ to
  $\approx50\%$ for more  absorbed systems. Poor statistics however,
  limit the  significance of the above result to the $\approx2\sigma$
  level. Also, the X-ray/radio matched sources have flatter coadded
  X-ray spectrum ($\Gamma=1.78^{+0.05}_{-0.03}$) compared to sources
  without radio emission ($\Gamma=2.00^{+0.03}_{-0.04}$). A possible
  explanation for the higher fraction of absorbed sources with radio
  emission at the $\rm \mu Jy$ level is the presence of circum-nuclear
  starburst activity that both feeds and obscures the central engine.
  For a small sub-sample of $z\approx0.4$ radio emitting AGNs with
  $\rm N_H>10^{22}\,cm^{-2}$ their combined spectrum exhibits a soft
  X-ray component that may be associated with star-formation activity,
  although other possibilities cannot be excluded.
  We also find that radio emitting AGNs make up about 13--20 per cent of
  the hard-band X-ray background depending on the adopted
  normalisation. 
\end{abstract}


\section{Introduction}\label{sec_intro}
A key development in X-ray astronomy with the advent of the {\it
Chandra} and the XMM-{\it Newton} missions has been the resolution of
the bulk of the hard-band X-ray background (XRB) into discrete sources
(e.g. Brandt et al. 2001; Giaconni et al. 2002). Elucidating the
nature of these sources however, remains an open issue and has been
the subject of  intensive observational studies over the last few
years. Multiwavelength follow-up programs suggest that the
hard X-ray selected population comprises a heterogeneous mix of
sources including broad line QSOs and Seyfert 1s, narrow emission-line
systems, passive galaxies with absorption line spectra and optically
faint systems with properties suggesting obscured AGN activity at high
redshifts (Barger et al. 2002; Mainieri et al. 2002; Fiore et
al. 2003; Piconcelli et al. 2003; Georgantopoulos et al. 2004).

Although the above studies have provided a wealth of information on
the nature of the sources that make up the bulk of the hard-band XRB
there is still limited information on their radio properties. At
bright flux limits, $f_X(\rm 2 - 10 \,keV) \approx 10^{-13} \, erg\,
s^{-1}\, cm^{-2}$, Akiyama et al. (2000) cross-correlated the ASCA
Large Sky Survey with the FIRST radio catalogue, finding an
identification rate of $\approx35$ per cent and a fraction of
radio-loud hard X-ray selected sources of about 10 per cent.  At
somewhat fainter fluxes, $f_X(\rm 5 - 10 \,keV) \approx 5 \times
10^{-14} \, erg\, s^{-1}\, cm^{-2}$, Ciliegi et al. (2003) used deeper
radio observations to explore the radio properties of the X-ray
sources detected in the HELLAS survey. They also find an
identification rate  of $\approx$30\ per cent much higher than that of
soft (0.5-2\,keV) X-ray selected samples, and argue that this is due
to both observational effects (e.g. deeper radio data) and the
hard-band selection. 

A high fraction ($\approx50$ per cent) of hard X-ray selected sources
with radio counterparts is also reported by Barger et al. (2001) using
deep {\it Chandra} ($f_X(\rm  2 - 10 \,keV) \approx 4 \times 10^{-15}
\, erg\, s^{-1}\, cm^{-2}$) and VLA ($S_{1.4}\rm \approx25\mu Jy$) 
data. Although these authors do not focus on the radio properties of 
hard X-ray selected sources, they show that  the X-ray/radio matched
population contributes as much as 26 per cent of the 2-8\,keV XRB.  
In the 0.5-8\,keV spectral band, Bauer et al. (2002) performed a
detailed study on the association between faint X-ray and radio
sources using the 1\,Ms {\it Chandra} dataset and ultra-deep VLA
observations.  These authors find that about one third of the X-ray
selected AGNs have radio counterparts and argue that the radio
emission at least in the  sub-sample of harder (i.e. obscured) AGNs is 
likely to be associated with circum-nuclear star-formation activity.
The largest overlap between X-ray and radio sources in the Bauer et
al. (2002) study is for the sub-sample of narrow emission line
galaxies likely to be star-forming  systems. The X-ray  sample used by 
these authors is however, selected in the 0.5-8\,keV spectral band and
is  therefore biased toward soft X-ray sources due to the higher
sensitivity of {\it Chandra} at soft energies.  

In this paper we further explore the nature of hard (2-8\,keV) X-ray
selected sources and their radio properties by combining an ultra-deep
and homogeneous radio survey, the Phoenix Deep Survey (PDS; Hopkins et
al.  2003), with a single 30\,arcmin diameter 50\,ksec XMM-{\it
  Newton} pointing. These observations are complemented by optical and
near-infrared photometry as well as optical spectroscopy. This
multiwavelength dataset is referred to here as the Phoenix/XMM-{\it
  Newton} survey (Georgakakis et al. 2003). The X-ray data reach a
limiting flux of $f_X(\rm 2 - 8 \,keV) \approx 7.7 \times 10^{-15} \,
erg\, s^{-1}\, cm^{-2}$, intermediate to brighter X-ray/radio matched
samples (e.g. Akiyama et al. 2000; Ciliegi et al. 2003) and deeper
surveys (e.g. Barger et al. 2001). The Phoenix/XMM-{\it Newton} survey
also has the advantage of larger areal coverage compared to {\it
Chandra} surveys, due to both the large XMM-{\it Newton} field-of-view
and the wide area and homogeneous radio observations of the
PDS. Moreover, in the present study we exploit the unparalleled
sensitivity of the XMM-{\it Newton} (5 times more effective area than
the {\it Chandra}) to explore the X-ray spectral properties of the
detected sources. 

Section \ref{phoenix} presents the multiwavelength data (optical,
near-infrared, radio and X-ray) available for the PDS, while Section
\ref{sample} details the sample used in the present study. Our results
are presented in Section \ref{results} and discussed in  Section
\ref{discussion}.  Finally, section \ref{conclusions} summarises our
conclusions. Throughout this paper we adopt $\rm \Omega_M=0.3$, $\rm
\Omega_\Lambda=0.7$ and $\rm H_{o}=70\,km\,s^{-1}\,Mpc^{-1}$.

\section{The Phoenix Deep Survey}\label{phoenix}
The Phoenix/XMM-{\it Newton} survey is a multiwavelength program combining
deep radio and X-ray observations with optical and near-infrared data
aiming to explore the association between the faint radio and X-ray
populations.  

\subsection{X-ray data}
The X-ray data consist of a single $50$\,ks pointing obtained
by the XMM-{\it Newton} on 2002 May 5 and centered at   
RA(J2000)=$01^{\rm h}12^{\rm m}52^{\rm s}$;
Dec.(J2000)=$-45^{\circ}33^{\prime}10.0^{\prime\prime}$. The EPIC
(European Photon Imaging Camera; Str\"uder et al. 2001; Turner et 
al. 2001) cameras were operated in full frame mode with the medium
filter applied. 

A full description of the data reduction can be found in Georgakakis
et al. (2003). In brief, the Science Analysis Software (SAS 5.3) was
used to produce event files for the PN and the two MOS
detectors. These were then screened for high particle  background
periods resulting in PN and MOS good time intervals of 39,444 and
41,273\,s respectively. To increase the signal--to--noise ratio and to
reach fainter fluxes the PN and the MOS event files have been combined
into a single event list using the {\sc merge} task of SAS. Images in
celestial coordinates with pixel size of 4.35\,arcsec were
extracted in the spectral bands 0.5--8\,keV (total), 0.5--2\,keV
(soft) and 2--8\,keV (hard) for both the merged and the individual PN
and MOS event files. We use the more sensitive (higher S/N ratio)
merged image for source detection and flux estimation, while the
individual PN and MOS images are used to calculate hardness
ratios. This is because the interpretation of hardness ratios is
simplified if the extracted count rates  are from one detector only. 

In this study we use the sources detected in the 2-8\,keV band
merged image using the {\sc ewavelet} task of SAS with a detection
threshold of $6\sigma$. These sources were visually inspected
and spurious detections clearly associated with CCD gaps, hot pixels
or lying close to the edge of the field of view were  removed. The
final catalogue comprises a  total of 43 X-ray sources to the limit
$f_X(\rm 2 - 8 \,keV) \approx 7.7 \times10^{-15} \,erg \,s^{-1}
\,cm^{-2}$.  

Count rates in the merged (PN+MOS) images as well as the
individual PN and MOS images are estimated within an 18\,arcsec
aperture. For the  background estimation we use the background maps
generated by the {\sc ewavelet} task of  SAS. A small
fraction of sources lie close to masked regions (CCD gaps or hot
pixels) on either the MOS or the PN detectors. This may introduce
errors in the estimated source  counts. To avoid this bias, the source
count rates (and hence the hardness ratios and the flux) are estimated
using the detector (MOS or PN) with no masked pixels in the vicinity
of the source.

To convert count rates to flux the Energy Conversion Factors
(ECF) of individual detectors are calculated assuming a power law
spectrum with $\Gamma=1.7$ and Galactic absorption $\rm N_H=2\times
10^{20} \rm {cm^{-2}}$ appropriate for the Phoenix field (Dickey \&
Lockman 1990). The mean ECF
for the mosaic of all three detectors is estimated by weighting the
ECFs of individual detectors by the respective exposure time.  For the
encircled energy correction, accounting for the energy fraction
outside the aperture within which source counts are accumulated, we
adopt the calibration  given by the XMM-{\it Newton} 
Calibration Documentation 
\footnote{http://xmm.vilspa.esa.es/external/xmm\_sw\_cal/calib \\
/documentation.shtml\#XRT}.

\subsection{Radio data}
The radio observations of the Phoenix Deep Survey (PDS\footnote{see
  also http://www.atnf.csiro.au/people/ahopkins/phoenix/}) were
carried out at the Australia Telescope Compact Array (ATCA) at
1.4\,GHz during several campaigns between 1994 and 2001 in the 6A, 6B
and 6C array configurations. The data cover a 4.56 square degree area
centered at RA(J2000)=$01^{\rm h}11^{\rm m}13^{\rm s}$
Dec.(J2000)=$-45\degr45\arcmin00\arcsec$, much larger than the
30\,arcmin diameter region covered by the XMM-{\it Newton}
observations. A detailed description of the radio observations, data
reduction and source detection are discussed by Hopkins et al. (1998,
1999, 2003). The observational strategy adopted resulted in a radio
map that is highly homogeneous within the central $\rm \approx1\,deg$
radius.  The $1\sigma$ rms noise nevertheless increases from $\rm
12\mu Jy$ at the most sensitive region to about $\rm 90\mu Jy$ close
to the field edge. The final catalogue consists of a total of 2148
radio sources to a limit of 60\,$\rm \mu$Jy (Hopkins et al.  2003).

The   XMM-{\it Newton} pointing lies within the most homogeneously
covered region of the PDS but is offset from the most sensitive area
of the radio map by about 0.30\,deg. Therefore, the completeness limit
of the  radio observations in that 30\,arcmin diameter  region is $\rm
\approx80\,\mu Jy$. The source detection was performed using
the  False Discovery Rate algorithm described by Hopkins et
al. (2002) to robustly quantify the number of spurious sources in
radio catalogues. This novel method is somewhat different from the
traditional  radio source detection algorithms that select
sources with peak flux density above a user defined threshold
expressed in multiples of the local background RMS
noise. Nevertheless, in the region  covered by the  XMM-{\it Newton}
the flux density limit of $\rm  \approx80\,\mu Jy$ corresponds to a
detection threshold for the traditional method of $\approx5\sigma$. A  
total of 204 radio sources overlap  with the XMM-{\it Newton} data.

\subsection{Optical and near-infrared photometry}
Optical photometric observations of most of the  PDS area in the $V$ and 
$R$-bands were obtained at the Anglo-Australian Telescope (AAT) during
two observing runs in  1994 and 1995. A detailed description
of these observations including data reduction, photometric
calibration, source detection and optical identification are
presented by Georgakakis et al. (1999). This dataset is
complete to $R=22.5$\,mag. The $U$-band data used in the present study
are from the ESO 2.2\,m telescope using the WFI on August 18
2001. These observations are described  in Sullivan et al. (2004) and
reach a completeness limit of $U\approx22.5$\,mag. 

Near-infrared observations of the 30\,arcmin diameter area covered
by the XMM-{\it Newton} pointing were obtained using the OSIRIS
multi-purpose instrument at the $f/13.5$ focus of the CTIO
1.5-m telescope. The observations were carried out in 2000 December
4-6 in the $J$ and  $Ks$ bands. The OSIRIS is equipped  with a
1024x1024 Hawaii HgCdTe array with a field of view of $\rm 11 \times
11 \, arcmin^{2}$ and a pixel scale of 1.153\,arcsec for the $f/2.8$
camera focus position used in our program. To cover the 30\,arcmin
diameter XMM-{\it Newton} pointing we used 7 contiguous pointings.   

In the $Ks$ band for each pointing a dithering pattern was adopted
that consisted of a sequence of 60\,s integrations followed by an
offset of the telescope. To get accurate sky frames the offset vector
was not replicated between successive exposures. The total integration
time varied between 20 to 40\,min with  a mean of 35\,min per
pointing. In the $J$-band a similar dithering procedure was followed
with an exposure time between offsets of 30 to 60\,s. The total
integration time was 20\,min per pointing.

The data reduction was carried out using {\sc iraf} tasks. The flat
field frame was constructed using both the target observations and the
dome flat following a method developed by Peter Witchalls and Will 
Saunders and described by Sullivan et al. (2004). The sky frame to
be subtracted from a given target frame was generated by median
combining the 10 images closest in time to the frame in question. All 
images of the same pointing were registered and coadded to produce the
final image. Photometric calibration was carried out using standard
stars from Persson et al. (1998). The photometric solutions were stable
from night-to-night with a zero-point variation of less than
0.1\,mag. The estimated uncertainty in the zero-point is less than
0.02\,mag. Astrometric calibration for each field was performed using
the positions of at least 20 USNO-2 stars. These solutions were
typically accurate to 0.5\,arcsec. Source detection used the {\sc
SExtractor} package (Bertin \& Arnouts 1996). Comparing the number
counts from these observations with previous studies we estimate a
completeness limit of $Ks\approx18$\,mag.  

Finally we note that the Phoenix/XMM-{\it Newton} field partly
overlaps with the deep optical $UBVRI$ photometric data presented by
Sullivan et al. (2004). These observations use the Wide Field Imager
at the AAT ($BVRI$ bands) and the CTIO-4m telescopes and reach
limiting magnitudes $R\approx24$ and $U\approx24$\,mag. When available
we use these deeper data rather than the shallower observations from
either the ESO 2.2m ($U$-band) or our previous AAT survey of the PDS
($VR$-bands). The magnitude given in this study are in the Vega
system.  

\subsection{Optical spectroscopy}\label{sec_spec}
A number of X-ray sources in the Phoenix/XMM-{\it Newton} field have
optical spectroscopic data from the 2dF facility at the AAT as part of
the on-going spectroscopic program aiming to obtain spectral
information for the optically identified faint radio sources
in the PDS. At  present redshifts and spectral classifications are
available for over 300 radio sources brighter than
$R=21.5$\,mag. This large dataset is presented by Georgakakis et
al. (1999) and Afonso et al. (2004 in preparation).  

Additional spectra for both radio and X-ray sources in the
Phoenix/XMM-{\it Newton} field were obtained with the HYDRA
multi-fibre spectrograph at the CTIO Blanco 4-m telescope. A 
detailed description of these observations as well as the full
spectroscopic catalogue will be presented  in a future paper
(Georgakakis et al. 2004 in preparation). In the present study we
concentrate on the HYDRA spectra of the hard X-ray selected sample. In
brief, the HYDRA has a 40\,arcmin diameter  field of view and is
equipped with 138 fibres, each 2\,arcsec in diameter. The
observations were  carried out in 2003   August 8-9 in photometric
conditions. The grating used was  the KPGL2 providing a dispersion of
$\rm 1.19\,\AA\,pixel^{-1}$ and a wavelength resolution of $\approx
\rm 8\,\AA$. To avoid second order contamination we used the GG420
blocking filter. As a result  the wavelength coverage of  
the KPGL2+GG420 combination is $\rm \approx4200-8300\,\AA$. 
The total exposure time for each source was 3.5 to 4\,h  
split into 5 or 6 half hour integrations.

The data were reduced using the {\sc hydra} reduction package within
{\sc iraf}. Fibre flat fields were used to trace the fibre spectra on
the CCD and to flat-field the data. Blank sky exposures taken between
target observations were employed for throughput calibration. CuArHe
arc lamp exposures were employed for wavelength calibration. Redshifts 
were determined  by visual inspection of the resulting spectra. These
were then classified into narrow emission line objects, absorption
line galaxies and broad emission line systems. Flux
calibration  has not been performed. This is due to the difficulty in
obtaining absolute flux calibration for fibres which can
differ substantially in their throughput. 

\section{The sample}\label{sample}
The hard (2-8\,keV) X-ray selected sample used in the present study
comprises a total of 43 sources to the limit  $f_X(\rm
2-8\,keV)\approx7.7\times10^{-15} \, erg\, s^{-1}\, cm^{-2}$. These
sources were optically identified using the $R$-band source catalogue
by  estimating the probability, $P$, a candidate counterpart is a
chance alignment (Downes et al. 1986). Counterpart candidates  were 
searched for within a 6\,arcsec radius. We propose 25 secure optical
identifications with $P<0.015$. We also find 5 sources with higher
but still acceptable probability being spurious coincidence,
$0.015<P<0.03$. 
One source optically identified in the much deeper $BVRI$ optical data 
presented by Sullivan et al. (2004)  lies $\approx2$\,arcsec  
from a very faint $R\approx24$\,mag  optical galaxy. Although the
probability that this source is a spurious alignment is $\approx6$ per
cent we accept the optical identification since the offset between the
X-ray and optical positions is within the XMM-{\it Newton}
positional uncertainty for X-ray bright sources ($\approx3.5$\,arcsec;  
McHardy et al. 2003). Summarising, of the 43 hard X-ray selected
sources 31 have optical identifications while  the remaining 12 are
blank fields to the limit $R\approx22.5$\,mag.   

The radio and the hard X-ray selected samples were positionally
matched using a radius of 6\,arcsec. We find a total of 7
sources in common between the two catalogues. We also search for lower
significance radio counterparts of hard X-ray selected sources by
looking for $>3\sigma$ peak radio emission in the vicinity ($\rm
<6\,arcsec$) of X-ray  sources. This method further identifies 7
X-ray/radio matches. Given 
the surface densities of the radio and the X-ray source catalogues the 
probability of finding by chance a $3\sigma$ radio source within
$\rm 6\,arcsec$ from an X-ray position is estimated to be
$\approx1$ per cent.

Redshift measurements are available for 18 hard X-ray selected sources
brighter than $R\approx22$\,mag. In addition to these spectroscopic
redshifts, for X-ray sources with optical and near-infrared (NIR)
photometry in at least 4 bands and relatively hard X-ray spectral
properties, $\rm HR>-0.4$ (corresponding to observed column densities
in excess of $\rm 10^{21}\, cm^{-2}$; $\Gamma=1.7$) we also estimate
photometric redshifts using galaxy templates and methods outlined in
Sullivan et al. (2004). These sources have colours suggesting
that their optical emission is dominated by light from the host galaxy
rather than the central AGN, thus allowing templates for normal
galaxies to be used (see section \ref{results}; Barger et al. 2002,
2003; Mobasher et al. 2004; Gandhi et al. 2004). Comparing $z_{phot}$
with $z_{spec}$ for the 5 sources with (i) $\rm HR>-0.4$, (ii) at
least 4-band photometry and (iii) available spectroscopy we estimate
the accuracy of our method to be $| \delta z|/z_{spec}\approx
0.1$. Although the statistics are poor the above result indicates that
templates for normal galaxies can successfully be used to estimate
photometric redshifts for the X-ray harder sources (see also  Barger
et al. 2002, 2003; Mobasher et al. 2004; Gandhi et al. 2004). This
method provides photometric redshift estimates for 3 sources that have
no spectroscopic redshift measurement. X-ray sources with soft X-ray
spectral properties are most likely dominated by light from the
central AGN (see section \ref{results}), although we caution the
reader that there is increasing evidence for broad line AGNs that do
not behave this way and show flat (hard) X-ray spectra (Risaliti et
al. 2001; Risaliti et al. 2003). Unlike galaxies, estimating
photometric  redshifts for AGN dominated systems is challenging and
despite recent progress the results are significantly less accurate
(Richards et al. 2001; Kitsionas et al. 2004; Babbedge et
al. 2004). In the present study we do not estimate photometric
redshifts for sources with soft X-ray spectral properties likely to be
Seyfert 1s or QSOs.  

One of the hard X-ray selected sources (\#25 in Table \ref{tbl1}
below) has a very bright, $R\approx13$\,mag, optically unresolved
counterpart and very low X-ray--to--optical flux ratio
($\log f_X/f_{opt}\approx-3$). Also, as discussed in Appendix
\ref{app1} the X-ray spectrum of this object is best fit by a
Raymond-Smith hot gas model (Raymond \& Smith 1977) with temperature
$\rm kT=0.7\,keV$. Although optical spectroscopy is not available the
evidence above indicates that this X-ray source is  associated with a
Galactic star.  In the rest of this paper we will not consider this
source in our analysis.     

The hard X-ray selected sample used in this paper is presented in
Table \ref{tbl1} which has the following format: 

{\bf 1.} Identification number. 

{\bf 2-3.} Right ascension and declination of the X-ray centroid
position in J2000. 

{\bf 4-7} $U$, $V$, $R$ and $K$-band magnitudes (Vega based system)
respectively of the optical counterpart if available.

{\bf 8.} Probability, $P$, the optical counterpart is a chance
coincidence. 

{\bf 9.} Offset in arcseconds between the X-ray and optical source
positions.

{\bf 10.} Spectroscopic or photometric redshift.  

{\bf 11.} Quality, $Q$,  of the redshift estimate. A value $Q=3$
corresponds to three or more identified spectral features indicating a
reliable redshift. A value $Q=1, 2$ corresponds to 1 and 2 identified
spectral  features respectively.  

{\bf 12.} Classification on the basis of the observed optical spectral
features: {\bf AB:}  absorption lines only; {\bf
NL:} narrow emission lines; {\bf BL:} broad emission lines.  

{\bf 13.} Radio flux density, $S_{1.4}$, of the radio counterpart of
the X-ray source if available. Radio sources that lie below the formal
limit $S_{1.4}\approx\rm 80\mu Jy$ of the radio catalogue corresponding
to about $5\sigma$ are marked. Although the peak flux density of these
sources is below the   $5\sigma$ level their integrated flux density
is, in some cases, brighter than  $\rm 80\mu Jy$. 

{\bf 14.} Offset in arcseconds between the radio and X-ray source
positions. 

\begin{table*}
\scriptsize
\begin{center}
\begin{tabular}{l ccc ccc ccc ccc c}
\hline
ID &
$\alpha_X$ &
$\delta_X$ &
$U$ &
$V$ &
$R$ &
$K$ &
$P$ &
$\delta_{XO}$ &
$z$    &
$Q$  &
class$^{a}$    &
$S_{\rm 1.4}$  &
$\delta_{XR}$ \\

  &
(J2000) &
(J2000) &
(mag) &
(mag) &
(mag) &
(mag) &
($\times10^{-3}$) &
($^{\prime\prime}$) &
       & 
       & 
       & 
(mJy)    &
($^{\prime\prime}$) \\
\hline

1 & 1 14 05.4 & -45 32 19 & $21.06\pm0.06$ & $-$ & $20.15\pm0.03$ & $17.56\pm0.12$ & 2.2 & 1.3 & 1.160 & 2 & BL & $-$ & $-$ \\

2 & 1 14 00.4 & -45 34 42 & $21.33\pm0.07$ & $-$ & $20.30\pm0.04$ & $17.50\pm0.12$ & 0.2 & 0.4 & 0.406 & 2 & NL & $-$ & $-$ \\

3 & 1 13 58.5 & -45 39 11 & $19.69\pm0.02$ & $-$ & $20.65\pm0.05$ & $>18$ & 17.1 & 3.0 & 1.385 & 3 & BL & $-$ & $-$ \\

4 & 1 13 49.8 & -45 27 02 & $21.83\pm0.09$ & $-$ & $20.49\pm0.04$ & $18.08\pm0.18$ & $<0.1$ & 0.1 & $-$ & $-$ & $-$ & $-$ & $-$ \\

5 & 1 13 39.8 & -45 32 31 & $21.13\pm0.05$ & $20.73\pm0.03$ & $-$ & $18.41\pm0.29$ & 0.5 & 0.6 & 2.329 & 3 & BL & $-$ & $-$ \\

6 & 1 13 37.6 & -45 31 45 & $>22.5$ & $22.55\pm0.13$ & $-$ & $>18$ & 20.5 & 2.3 & $-$ & $-$ & $-$ & $-$ & $-$ \\

7 & 1 13 33.2 & -45 38 42 & $20.52\pm0.06$ & $19.52\pm0.02$ & $18.49\pm0.01$ & $15.72\pm0.06$ & 0.4 & 0.9 & 0.276 & 3 & NL & 0.472 & 1.4 \\

8 & 1 13 31.9 & -45 36 28 & $22.89\pm0.23$ & $22.04\pm0.12$ & $21.1\pm0.05$ & $17.90\pm0.16$ & 26.5 & 3.2 & 0.516 & 1 & NL & $-$ & $-$ \\

9 & 1 13 31.7 & -45 44 43 & $>22.5$ & $>23$ & $>22.5$ & $>18$ & $-$ & $-$ & $-$ & $-$ & $-$ & $-$ & $-$ \\

10& 1 13 27.0 & -45 38 32 & $19.73\pm0.02$ & $19.99\pm0.02$ & $19.62\pm0.01$ & $17.84\pm0.28$ & 4.5 & 2.2 & 1.220 & 2 & BL & $-$ & $-$ \\

11& 1 13 19.9 & -45 32 58 & $>22.5$ & $>23$ & $>22.5$ & $>18$ & $-$ & $-$ & $-$ & $-$ & $-$ & $-$ & $-$ \\

12& 1 13 14.6 & -45 40 52 & $22.32\pm0.19$ & $22.09\pm0.14$ & $22.03\pm0.15$ & $>18$ & 4.9 & 1.1 & 0.985 & 1 & NL & 1.027 & 1.8 \\

13& 1 13 11.5 & -45 24 01 & $20.48\pm0.03$ & $20.12\pm0.02$ & $19.46\pm0.01$ & $16.68\pm0.07$ & 0.5 & 0.8 & 0.299 & 3 & BL & 0.125$^{e}$ & 4.7 \\

14& 1 13 10.7 & -45 38 20 & $21.90\pm0.10$ & $21.41\pm0.06$ & $21.34\pm0.06$ & $>18$ & 9.5 & 1.9 & $-$ & $-$ & $-$ & 0.039$^{e}$ & 3.5 \\

15& 1 13 09.9 & -45 41 37 & $>22.5$ & $>23$ & $>22.5$ & $>18$ & $-$ & $-$ & $-$ & $-$ & $-$ & $-$ & $-$ \\

16& 1 13 09.9 & -45 32 30 & $>22.5$ & $>23$ & $>22.5$ & $>18$ & $-$ & $-$ & $-$ & $-$ & $-$ & $-$ & $-$ \\

17& 1 13 06.3 & -45 45 18 & $19.98\pm0.04$ & $18.45\pm0.01$ & $17.67\pm0.03$ & $14.58\pm0.03$ & 6.1 & 4.7 & 0.235 & 3 & AB & $-$ & $-$ \\

18& 1 13 02.8 & -45 24 41 & $>22.5$ & $22.55\pm0.13$ & $-$ & $>18$ & 18.0 & 1.3 & $-$ & $-$ & $-$ & $-$ & $-$ \\

19& 1 13 02.0 & -45 30 22 & $18.85\pm0.01$ & $19.54\pm0.01$ & $19.22\pm0.01$ & $17.83\pm0.15$ & 0.2 & 0.7 & 1.208 & 1 & BL & $-$ & $-$ \\

20& 1 13 01.1 & -45 38 47 & $>22.5$ & $>23$ & $>22.5$ & $18.65\pm0.38$ & $-$ & $-$ & $-$ & $-$ & $-$ & 0.063$^e$ & 1.2 \\

21& 1 13 00.0 & -45 37 17 & $>22.5$ & $-$ & $22.34\pm0.16$ & $18.00\pm0.25$ & 4.2 & 0.6 & $-$ & $-$ & $-$ & $-$ & $-$ \\

22& 1 12 59.1 & -45 23 43 & $>22.5$ & $20.98\pm0.06$ & $20.00\pm0.02$ & $17.08\pm0.09$ & 1.5 & 1.0 & 0.369 & 3 & NL & 0.078$^e$ & 2.4 \\

23$^{b}$& 1 12 58.8 & -45 38 22 & $18.98\pm0.01$ & $19.89\pm0.01$ & $19.75\pm0.02$ & $17.87\pm0.13$ & 2.2 & 1.5 & 1.975 & 3 & BL & 0.420$^e$ & 4.9 \\

24& 1 12 57.0 & -45 29 43 & $22.38\pm0.17$ & $22.19\pm0.15$ & $21.78\pm0.09$ & $>18$ & 10.9 & 1.7 & 1.550 & 1 & BL & $-$ & $-$ \\

25$^{c}$& 1 12 51.7 & -45 23 54 & $15.57\pm0.01$ & $13.88\pm0.01$ & $13.07\pm0.01$ & $11.01\pm0.02$ & 0.1 & 2.0 & $-$ & $-$ & STAR? & $-$ & $-$ \\

26& 1 12 49.9 & -45 32 55 & $>22.5$ & $>23$ & $>22.5$ & $>18$ & $-$ & $-$ & $-$ & $-$ & $-$ & $-$ & $-$ \\

27& 1 12 46.2 & -45 40 15 & $>22.5$ & $>23$ & $>22.5$ & $>18$ & $-$ & $-$ & $-$ & $-$ & $-$ & $-$ & $-$ \\

28& 1 12 43.8 & -45 42 50 & $22.29\pm0.07$ & $22.75\pm0.13$ & $21.60\pm0.01$ & $17.69\pm0.16$ & 1.7 & 2.1 & 0.776 & $-$ & photo-z & 0.062$^e$ & 2.3 \\

29$^{b}$& 1 12 43.7 & -45 35 45 & $23.72\pm0.15$ & $23.84\pm0.19$ & $23.41\pm0.03$ & $>18$ & 23.0 & 1.5 & $-$ & $-$ & $-$ & 0.077 & 0.1 \\

30& 1 12 42.2 & -45 28 39 & $20.58\pm0.03$ & $21.15\pm0.06$ & $20.85\pm0.04$ & $18.79\pm0.45$ & 4.4 & 1.5 & $-$ & $-$ & $-$ & -- & -- \\

31& 1 12 40.5 & -45 38 41 & $>22.5$ & $>23$ & $>22.5$ & $>18$ & $-$ & $-$ & $-$ & $-$ & $-$ & 0.137 & 0.5 \\

32& 1 12 32.3 & -45 35 20 & $20.51\pm0.03$ & $20.58\pm0.04$ & $19.95\pm0.02$ & $>18$ & 2.0 & 1.2 & $-$ & $-$ & $-$ & 0.070$^{e}$ & 0.3 \\

33& 1 12 28.0 & -45 26 41 & $>22.5$ & $22.55\pm0.13$ & $-$ & $>18$ & $-$ & $-$ & $-$ & $-$ & $-$ & 0.121 & 2.1 \\

34$^{b}$& 1 12 27.1 & -45 44 27 & $20.64\pm0.01$ & $21.30\pm0.01$ & $21.09\pm0.01$ & $18.43\pm0.40$ & 6.5 & 1.9 & 1.010 & 2 & BL & $-$ & $-$ \\

35& 1 12 23.1 & -45 24 56 & $>22.5$ & $>23$ & $>22.5$ & $-$ & $-$ & $-$ & $-$ & $-$ & $-$ & $-$ & $-$ \\

36$^{b}$& 1 12 15.0 & -45 31 00 & $23.15\pm0.01$ & $22.39\pm0.04$ & $21.34\pm0.01$ & $17.7\pm0.13$ & 3.9 & 1.0 & 0.842 & $-$ & photo-z & $-$ & $-$ \\

37$^{b}$& 1 12 13.8 & -45 45 52 & $19.39\pm0.01$ & $19.80\pm0.01$ & $19.67\pm0.01$ & $-$ & 2.5 & 1.6 & 0.922 & 3 & BL & $-$ & $-$ \\

38& 1 12 12.5 & -45 27 47 & $>22.5$ & $>23$ & $>22.5$ & $-$ & $-$ & $-$ & $-$ & $-$ & $-$ & 0.181 & 2.8 \\

39& 1 12 10.1 & -45 30 05 & $>22.5$ & $>23$ & $>22.5$ & $>18$ & $-$ & $-$ & $-$ & $-$ & $-$ & $-$ & $-$ \\

40$^{bd}$& 1 12 04.8 & -45 35 32 & $>25$ & $24.80\pm0.30$ & $23.85\pm0.10$ & $>18$ & 62.5 & 2.0 & 0.732 & $-$ & photo-z & $-$ & $-$ \\

41& 1 12 04.6 & -45 25 49 & $21.16\pm0.03$ & $21.55\pm0.06$ & $21.27\pm0.04$ & $-$ & 3.8 & 1.2 & 1.200 & 1 & BL & $-$ & $-$ \\

42$^{b}$& 1 11 50.2 & -45 37 25 & $21.25\pm0.03$ & $20.37\pm0.01$ & $19.46\pm0.01$ & $-$ & 2.8 & 2.1 & 0.493 & 3 & NL & 0.135 & 2.1 \\

43$^{b}$& 1 11 41.9 & -45 40 01 & $20.99\pm0.01$ & $20.46\pm0.01$ & $19.78\pm0.01$ & $-$ & 3.0 & 1.8 & $-$ & $-$ & $-$ & $-$ & $-$ \\
\hline
\multicolumn{14}{l}{$^a$AB: absorption lines; NL: Narrow emission lines;
BL: Broad emission lines; photo-z: photometric redshift} \\
\multicolumn{14}{l}{$^b$UV and optical data are from Sullivan et al. (2004)} \\
\multicolumn{14}{l}{$^c$Optical and X-ray properties indicating Galactic star} \\
\multicolumn{14}{l}{$^d$source \#40: although $P>0.05$ the optical ID is included in the sample since 
$\delta_{OX}$ is within the XMM-{\it Newton} positional uncertainty.}\\  
\multicolumn{14}{l}{$^e$lower significance radio source with peak flux
density in the range $3-5\sigma$. Not included in the Hopkins et
al. (2003) radio catalogue.}
\end{tabular}
\end{center}
\caption{
Phoenix/XMM-{\it Newton} survey: optical and radio properties of the hard X-ray selected sample.
}\label{tbl1}
\normalsize
\end{table*}

We further explore the X-ray spectral properties of the present sample
using  the {\sc xspec} v11.2 package. For sources with small number of
net counts we use the C-statistic technique  (Cash 1979) specifically
developed to extract information from low signal-to-noise ratio
spectra. The data are grouped to have at least one count per
bin. We note however, that higher binning factors or no binning does
not change our results.  Firstly, we attempt to constrain the $\rm
N_H$ by fitting an absorbed (Wisconsin cross-sections; Morrison and
McCammon 1983) power-law model (wabs*pow) fixing the power-law index
to $\Gamma=1.7$. This value of $\Gamma$ is selected to be  inbetween
the mean spectral index of radio loud ($\Gamma=1.6$; Reeves \& Turner
2000; Gambill 2003) and radio quiet AGNs ($\Gamma\approx1.9$;
Laor et al. 1997; Reeves \& Turner 2000). We then use the same model
(wabs*pow) to estimate the power law index $\Gamma$ keeping the column
density fixed to the Galactic value ($\rm
N_H=2\times10^{20}\,cm^{-2}$). For sources with sufficient counts we
perform standard $\chi^{2}$ spectral fitting. The data were  grouped
to have a minimum of 15 counts per bin to ensure that Gaussian
statistics apply. For the  $\chi^{2}$ analysis  we require that the
source spectrum has at least 15 spectral bins. An absorbed power-law
(wabs*pow) is fit to the data yielding the intrinsic absorbing
column density (i.e. after subtracting the Galactic absorption) and
the power-law photon index $\Gamma$. This model provides acceptable
fits  (i.e. reduced $\chi^{2}\approx1$) for all sources. The
parameters estimated from the C-statistic and the $\chi^{2}$ analysis
are consistent within the errors.  
For both the $\chi^{2}$ and the  C-statistic analysis the fit was
performed in the 0.2-8\,keV energy range where the sensitivity of the
XMM-{\it Newton} is the highest. The estimated errors correspond to
the 90 per cent confidence level.  The results of the above X-ray
spectral analysis along with the X-ray properties of the sample are
presented in Table \ref{tbl2} which has the following format:      

{\bf 1.} Identification number. 

{\bf 2.}  2-8\,keV X-ray flux in $\rm erg\,s^{-1}\,cm^{-2}$. 

{\bf 3.} Hardness ratio, $\rm HR$, defined as
\begin{equation}\label{eq1}
\rm HR = \frac{\rm RATE(2080)-RATE(0520)}{\rm RATE(2080)+RATE(0520)},
\end{equation}
where $\rm RATE(0520)$ and $\rm RATE(2080)$ are the count rates in
the 0.5-2 and 2-8\,keV spectral bands respectively. For sources with
less than 5 net counts in either the  hard or the soft  bands a lower
or an upper limit ($3\sigma$) respectively is estimated for the
hardness ratio assuming Poisson statistics. The hardness ratios are
estimated using the PN data except for sources that lie close
to PN CCD gaps or hot pixels where we use MOS data (see section
2.1). These sources are marked in Table \ref{tbl2}.  

{\bf 4.} Column density $\rm N_H$ estimated by either the C-statistic
method assuming $\Gamma=1.7$ or the standard $\chi^{2}$  spectral
fitting in the case of sources with sufficient counts (see discussion
above).  

{\bf 5.} Power law spectral index  $\rm \Gamma$ estimated by either the
C-statistic for a fixed Galactic column density $\rm
N_H=2\times10^{20}\,cm^{-2}$  or the standard $\chi^{2}$  spectral
fitting in the case of sources with sufficient counts (see discussion
above).  

{\bf 6.} 2-8\,keV X-ray luminosity if a spectroscopic or photometric 
redshift is available. For the k-correction we  assume $\Gamma=1.7$.

{\bf 7.} 1.4\,GHz radio power  if a spectroscopic or photometric 
redshift is available, adopting a radio spectral index $\alpha=0.8$
for the k-correction.

\begin{table*}
\footnotesize
\begin{center}
\begin{tabular}{l ccc ccc}
\hline
ID &
$f_X(\rm 2-8\,keV)$ &
HR &
$\rm N_{H}$$^{a}$ &
$\Gamma$ &
$\log L_X(\rm 2-8\,keV)$ &
$\log L_{1.4}$ \\

  & 
$(\times 10^{-14} \rm erg\,s^{-1}\,cm^{-2})$ &
  & 
($\rm 10^{21}\,cm^{-2}$) &
 & 
$(\rm erg\,s^{-1})$ &
($\rm W\,Hz^{-1}$) \\

\hline
1$^{}$ & $3.39\pm0.46$ & $-0.51\pm0.09$ & $<0.52$ & $1.99_{-0.26}^{+0.35}$ & $44.40\pm3.41$ & $-$ \\

2$^{b}$ & $1.47\pm0.48$ & $-0.70\pm0.10$ & $<0.19$ & $2.69_{-0.27}^{+0.27}$ & $42.97\pm1.36$ & $-$ \\

3$^{}$ & $2.87\pm0.47$ & $-0.36\pm0.11$ & $<0.19$ & $2.02_{-0.23}^{+0.27}$ & $44.51\pm2.86$ & $-$ \\

4$^{bc}$ & $0.89\pm0.41$ & $-0.59\pm0.17$ & $<0.09$ & $2.42_{-0.32}^{+0.26}$ & $-$ & $-$ \\

5$^{}$ & $1.60\pm0.30$ & $-0.66\pm0.09$ & $<0.56$ & $1.94_{-0.21}^{+0.42}$ & $44.78\pm2.55$ & $-$ \\

6$^{c}$ & $1.23\pm0.28$ & $-0.28\pm0.18$ & $2.70_{-1.00}^{+2.28}$ & $1.23_{-0.24}^{+0.22}$ & $-$ & $-$ \\

7$^{c}$ & $6.66\pm0.51$ & $+0.49\pm0.09$ & $21.79_{-4.39}^{+0.91}$ & $0.33_{-0.24}^{+0.12}$ & $43.25\pm5.76$ & 23.21 \\

8$^{bc}$ & $1.07\pm0.36$ & $-0.14\pm0.22$ & $7.06_{-1.76}^{+10.84}$ & $0.84_{-0.43}^{+0.18}$ & $43.08\pm1.32$ & $-$ \\

9$^{c}$ & $3.08\pm0.50$ & $-0.03\pm0.17$ & $13.30_{-4.80}^{+9.30}$ & $0.89_{-0.49}^{+0.36}$ & $-$ & $-$ \\

10$^{}$ & $2.34\pm0.43$ & $-0.61\pm0.07$ & $0.20_{-0.20}^{+0.39}$ & $2.35_{-0.25}^{+0.31}$ & $44.29\pm2.55$ & $-$ \\

11$^{}$ & $1.29\pm0.23$ & $-0.50\pm0.13$ & $<1.40$ & $1.67_{-0.27}^{+0.60}$ & $-$ & $-$ \\

12$^{c}$ & $2.10\pm0.35$ & $+0.51\pm0.24$ & $22.4_{-7.90}^{+13.30}$ & $0.36_{-0.36}^{+0.35}$ & $44.03\pm2.75$ & 25.01 \\

13$^{}$ & $1.28\pm0.31$ & $-0.71\pm0.09$ & $<0.5$ & $2.54_{-0.17}^{+0.37}$ & $42.61\pm1.82$ & 22.71 \\

14$^{}$ & $1.62\pm0.27$ & $-0.55\pm0.11$ & $0.11_{-0.11}^{+0.70}$ & $1.74_{-0.24}^{+0.35}$ & $-$ & $-$ \\

15$^{c}$ & $1.55\pm0.34$ & $+0.30\pm0.30$ & $10.20_{-3.60}^{+7.20}$ & $0.43_{-0.36}^{+0.41}$ & $-$ & $-$ \\

16$^{b}$ & $1.21\pm0.28$ & $-0.43\pm0.11$ & $<1.40$ & $2.02_{-0.28}^{+0.58}$ & $-$ & $-$ \\

17$^{}$ & $2.96\pm0.51$ & $-0.41\pm0.12$ & $<0.97$ & $1.58_{-0.27}^{+0.54}$ & $42.73\pm2.57$ & $-$ \\

18$^{}$ & $2.00\pm0.32$ & $-0.56\pm0.10$ & $0.70_{-0.70}^{+0.90}$ & $2.00_{-0.22}^{+0.49}$ & $-$ & $-$ \\

19$^{}$ & $1.59\pm0.22$ & $-0.86\pm0.05$ & $<0.04$ & $3.27_{-0.14}^{+0.07}$ & $44.12\pm3.32$ & $-$ \\

20$^{c}$ & $0.95\pm0.25$ & $+0.32\pm0.43$ & $51.67_{-23.47}^{+39.53}$ & $-0.25_{-0.87}^{+0.66}$ & $-$ & $-$ \\

21$^{b}$ & $1.13\pm0.31$ & $-0.65\pm0.10$ & $0.84_{-0.84}^{+0.96}$ & $2.24_{-0.49}^{+0.53}$ & $-$ & $-$ \\

22$^{c}$ & $1.28\pm0.31$ & $>0.13$ & $68.4_{-30.90}^{+59.60}$ & $-0.63_{-0.88}^{+0.91}$ & $42.81\pm1.83$ & 22.73 \\

23$^{}$ & $0.94\pm0.23$ & $-0.48\pm0.16$ & $<0.16$ & $1.58_{-0.43}^{+0.65}$ & $44.39\pm1.95$ & 25.51 \\

24$^{}$ & $0.79\pm0.19$ & $-0.46\pm0.13$ & $<0.28$ & $2.20_{-0.32}^{+0.34}$ & $44.07\pm1.97$ & $-$ \\

25$^{d}$ & $1.73\pm0.34$ & $-0.91\pm0.04$ & -- & -- & $-$ & $-$ \\

26$^{c}$ & $0.27\pm0.15$ & $-0.38\pm0.44$ & $0.55_{-0.55}^{+0.75}$ & $1.62_{-0.38}^{+0.40}$ & $-$ & $-$ \\

27$^{}$ & $1.44\pm0.30$ & $-0.51\pm0.12$ & $0.29_{-0.29}^{+1.41}$ & $1.89_{-0.30}^{+0.58}$ & $-$ & $-$ \\

28$^{c}$ & $1.00\pm0.40$ & $+0.22\pm0.33$ & $8.20_{-3.60}^{+5.20}$ & $0.91_{-0.37}^{+0.33}$ & $43.46\pm1.14$ & 23.50 \\

29$^{bc}$ & $0.91\pm0.27$ & $+0.08\pm0.25$ & $<0.10$ & $2.09_{-0.33}^{+0.32}$ & $-$ & $-$ \\

30$^{}$ & $1.22\pm0.22$ & $-0.54\pm0.10$ & $<0.33$ & $2.11_{-0.22}^{+0.26}$ & $-$ & $-$ \\

31$^{}$ & $2.35\pm0.36$ & $-0.18\pm0.10$ & $1.20_{-1.20}^{+3.50}$ & $1.59_{-0.32}^{+0.57}$ & $-$ & $-$ \\

32$^{}$ & $12.4\pm0.51$ & $-0.60\pm0.02$ & $<0.03$ & $2.10_{-0.04}^{+0.05}$ & $-$ & $-$ \\

33$^{c}$ & $1.70\pm0.29$ & $-0.26\pm0.16$ & $4.30_{-2.50}^{+1.50}$ & $1.02_{-0.20}^{+0.10}$ & $-$ & $-$ \\

34$^{}$ & $3.37\pm0.57$ & $-0.62\pm0.09$ & $0.27_{-0.27}^{+0.64}$ & $2.00_{-0.40}^{+0.21}$ & $44.26\pm2.71$ & $-$ \\

35 $^{bc}$ & $3.20\pm0.53$ & $+0.41\pm0.15$ & $14.28_{-3.29}^{+6.51}$ & $0.39_{-0.26}^{+0.26}$ & $-$ & $-$ \\

36$^{c}$ & $0.80\pm0.23$ & $+0.19\pm0.53$ & $9.90_{-4.50}^{+12.20}$ & $0.75_{-0.44}^{+0.42}$ & $43.45\pm1.54$ & $-$ \\

37$^{}$ & $4.95\pm0.83$ & $-0.46\pm0.08$ & $<0.29$ & $1.96_{-0.17}^{+0.21}$ & $44.33\pm2.73$ & $-$ \\

38$^{c}$ & $0.94\pm0.28$ & $-0.13\pm0.24$ & $2.34_{-2.27}^{+3.36}$ & $1.38_{-0.50}^{+0.50}$ & $-$ & $-$ \\

39$^{c}$ & $0.76\pm0.26$ & $+0.13\pm0.56$ & $5.50_{-3.50}^{+7.70}$ & $0.99_{-0.55}^{+0.49}$ & $-$ & $-$ \\

40$^{c}$ & $1.19\pm0.3$ & $-0.36\pm0.26$ & $4.3_{-1.9}^{+-3.6}$ & $0.98_{-0.28}^{+0.27}$ & $43.48\pm1.78$ & $-$ \\

41$^{}$ & $1.85\pm0.37$ & $-0.59\pm0.11$ & $<0.78$ & $2.00_{-0.22}^{+0.49}$ & $44.18\pm2.31$ & $-$ \\

42$^{c}$ & $1.45\pm0.40$ & $>-0.24$ & $68.7_{-37.2}^{+68.6}$ & $-0.27_{-0.71}^{+0.68}$ & $43.16\pm1.6$ & 23.3 \\

43$^{}$ & $5.81\pm0.67$ & $-0.48\pm0.08$ & $0.19_{-0.19}^{+0.59}$ & $2.04_{-0.27}^{+0.35}$ & $-$ & $-$ \\
\hline
\multicolumn{7}{l}{$^a$Observed column density not corrected for redshift} \\
\multicolumn{7}{l}{$^b$flux and hardness ratio from the MOS detector} \\
\multicolumn{7}{l}{$^c$$\rm N_H$ and $\Gamma$ estimated using the C-statistic method.} \\
\multicolumn{7}{l}{$^d$Galactic star candidate. Best fit by a Raymond-Smith model, kT=0.7, abundance=0.04} \\

\end{tabular}
\end{center}
\caption{
Phoenix/XMM-{\it Newton} survey: The X-ray properties, X-ray and radio luminosities of the sample
}\label{tbl2}
\normalsize
\end{table*}

Summarising the information presented in Tables \ref{tbl1} and
\ref{tbl2} the hard X-ray selected sample comprises: (i) 11
AGNs with broad emission lines,  (ii) 6 systems with narrow emission
line optical spectra, (iii) 1 galaxy with absorption  optical lines,
(iv) 1 Galactic star candidate on the basis of its optical and X-ray
properties, (v) 3 sources with photometric redshift estimates, (vi)  9
sources with optical identification but no redshift information most
of which have soft X-ray spectral properties suggesting AGNs and (vii)
12  optically unidentified sources. For the X-ray/radio matched  
population 4 sources are associated with narrow emission line systems,
2 show broad emission lines and 8 sources with no 
spectroscopic identification. Notes on individual optically identified
X-ray sources with radio counterparts and available spectroscopic
information are presented in Appendix \ref{app1}. This appendix
includes the optical and X-ray spectra as well as the optical images,
with  the X-ray contours overlaid,  of individual radio sources
(Figures  \ref{fig_spec}, \ref{fig_image} and
\ref{fig_xspec_app}). Also most of   the sources with radio
counterparts in the present study lie in the border line between
radio-loud and quiet systems ($\alpha_{RO}\approx0.35$; Stocke et
al. 1991; Ciliegi et al. 2003).          

We note that source \#12  in Tables \ref{tbl1} and \ref{tbl2} is
assigned a redshift $z=0.985$ on the basis of a single narrow emission
line interpreted as $\rm  [O\,II]\,3727\,\AA$.  The X-ray spectrum
suggests heavy obscuration corresponding to a rest-frame column
density (i.e. after correcting for redshift) of $\rm N_H\approx10^{23}
\,cm^{-2}$ ($\Gamma=1.7$). The X-ray luminosity of this source at
$z=0.985$ before applying any correction due to intrinsic
photoelectric  absorption is estimated  $L_X(\rm 2 - 8
\,keV)\approx10^{44} \, erg \, s^{-1} \, cm^{2}$. The evidence above
suggests a candidate type-2 QSO, although more observations are
required to confirm the redshift measurements and to search for broad
emission line components. More details on the properties of this
source are given in Appendix \ref{app1}.  

\begin{figure} 
\centerline{\psfig{figure=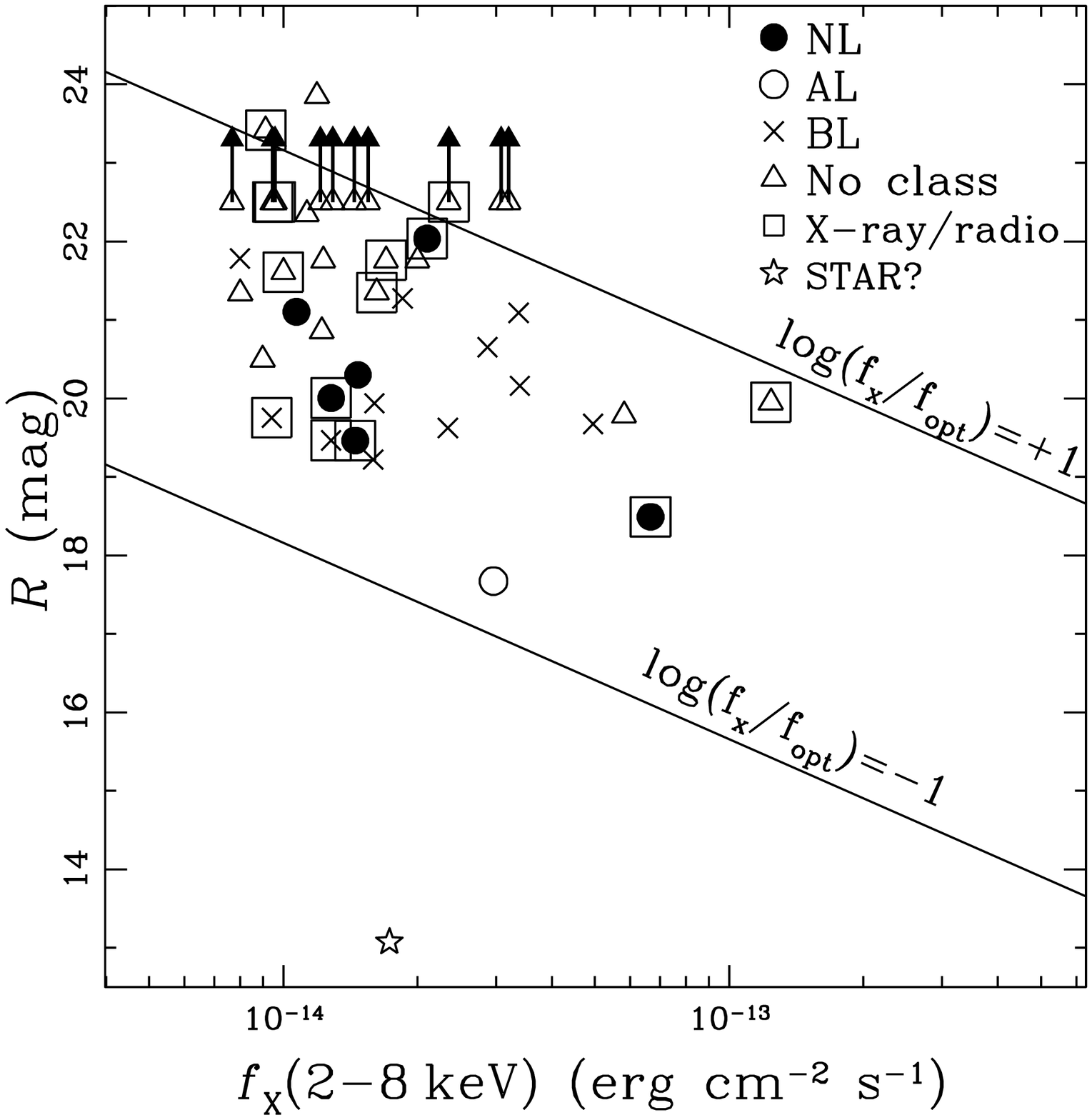,width=3.5in,angle=0}} 
\caption
 {$R$-band magnitude against 2-8\,keV flux. Filled circles are for
 sources with narrow emission line optical spectra, open circles
 correspond to absorption line systems, while crosses signify broad
 optical emission lines. Sources with no spectroscopic
 classification are shown with a triangle. A square on top of a symbol
 is for X-ray/radio matched sources. Optically unidentified objects as
 plotted as upper limits. The star is for the Galactic star candidate
 in the present sample. The lines indicate constant X-ray--to--optical
 flux ratios of +1 and --1. The lines $\log f_X/f_{opt}=\pm1$
 delineate the region of the parameter space occupied by powerful
 unobscured AGNs. 
 }\label{fig_fxfo}    
\end{figure}

\section{Results}\label{results}

\subsection{X-ray and optical properties of the sample}

Figure \ref{fig_fxfo} plots $R$-band magnitude against 2-8\,keV  
X-ray flux. The  $\log (f_X/f_{opt})=\pm1$ lines in this figure
delineate the region of the parameter space occupied by powerful
 unobscured AGNs (Stocke et al. 1991). The X-ray--to--optical flux
ratio is estimated from the relation  
\begin{equation}\label{eq2}
\log\frac{f_X}{f_{opt}} = \log f_X(2-8\,{\rm keV}) +
0.4\,R +5.53.
\end{equation}
The equation above is derived from the X-ray--to--optical flux
ratio definition of Stocke et al. (1991) that involved 0.3-3.5\,keV
flux and $V$-band magnitude. These quantities are converted to
2-8\,keV flux and $R$-band magnitude assuming a mean colour
$V-R=0.7$ and a power-law X-ray spectral energy distribution with
index  $\Gamma=1.7$. In Figure \ref{fig_fxfo} all the extragalactic
hard X-ray selected sources lie in the AGN region of the parameter
space (i.e. between the $\log f_X/f_{opt}\pm1$ diagonal lines). Hard
X-ray sources with radio counterparts have a range of
X-ray--to--optical flux ratios. 

\begin{figure}
\centerline{\psfig{figure=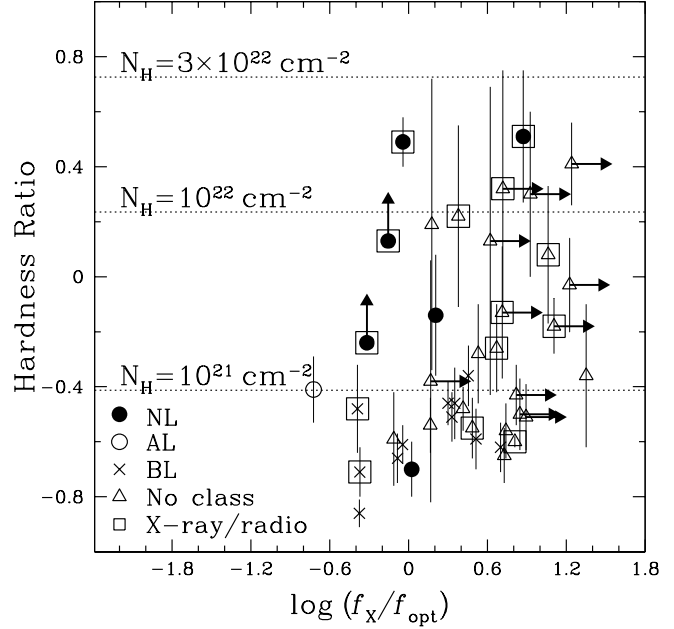,width=3.5in,angle=0}} 
\caption
 {Hardness ratio against X-ray--to--optical flux ratio as defined in
 section equation \ref{eq2}.  The symbols are the same as in
 Figure \ref{fig_fxfo}.  Sources that have less than 5 net counts in
 either the hard or the soft spectral bands are plotted as lower or
 upper limits respectively. The horizontal lines are the expected
 hardness ratio of a power-law spectral energy distribution with
 $\Gamma=1.7$ and different absorbing column densities in the range
 0.1 to $\rm 3\times10^{22}\,cm^{-2}$.
 }\label{fig_hr_vs_fxfo}    
\end{figure}

Optical unidentified sources  are shown as upper limits in this
figure. Some of them lie above the upper  bound of the empirical AGN 
envelope defined by  Stocke et al. (1991) suggesting
high redshift and/or dust obscuration (Alexander  et al. 2001; Brusa
et al. 2003; Fiore et al. 2003; Gandhi et al. 2004; Mignoli et
al. 2004; Georgantopoulos et al. 2004). Although  none of  these
sources  is  detected in our relatively shallow $K$-band survey  
($K=18$\,mag) they have, on average, hard X-ray spectral properties.
This is shown in Figure \ref{fig_hr_vs_fxfo} plotting hardness ratio
(equation \ref{eq1}) as a function of the X-ray--to--optical flux
ratio (equation \ref{eq2}). Optically unidentified sources with $\log  
f_X/f_{opt}\ga +1$ have $\rm HR>-0.2$ suggesting enhanced observed
photoelectric absorption  ($\rm N_H>3\times10^{21}\, cm^{-2}$,  
$\Gamma=1.7$). Also, in Figure \ref{fig_hr_vs_fxfo} there is fair
agreement between the X-ray and optical spectroscopic properties of
our sources. Spectroscopically confirmed broad-line AGNs have soft  
X-ray spectra, while narrow emission-line systems have,  
on average,  HRs suggesting observed absorbing columns in excess of
$\rm 10^{21}\,cm^{-2}$ ($\Gamma=1.7$). 

\begin{figure} 
\centerline{\psfig{figure=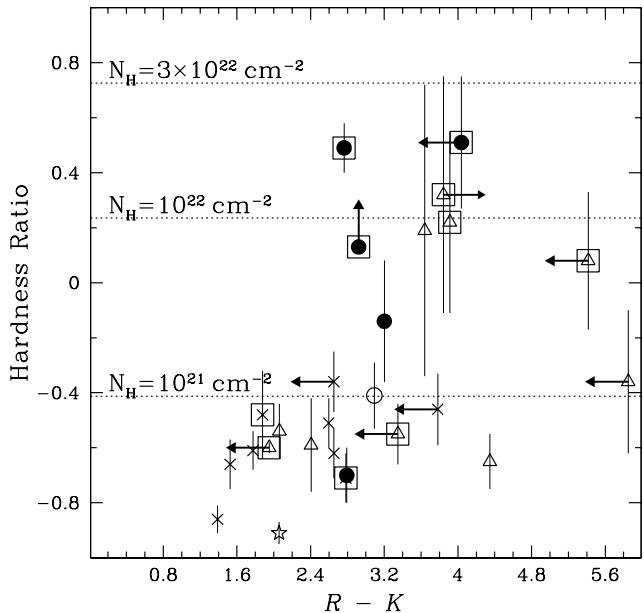,width=3.5in,angle=0}} 
\caption
 {Hardness ratio against $R-K$ colour. The symbols are the same as in
 Figure \ref{fig_fxfo}. The horizontal lines are the expected
 hardness ratio of a power-law spectral energy distribution with
 $\Gamma=1.7$ and different absorbing column densities in the range
 0.1 to $\rm 3\times10^{22}\,cm^{-2}$.  
 }\label{fig_hr_rk}    
\end{figure}

\begin{figure} 
\centerline{\psfig{figure=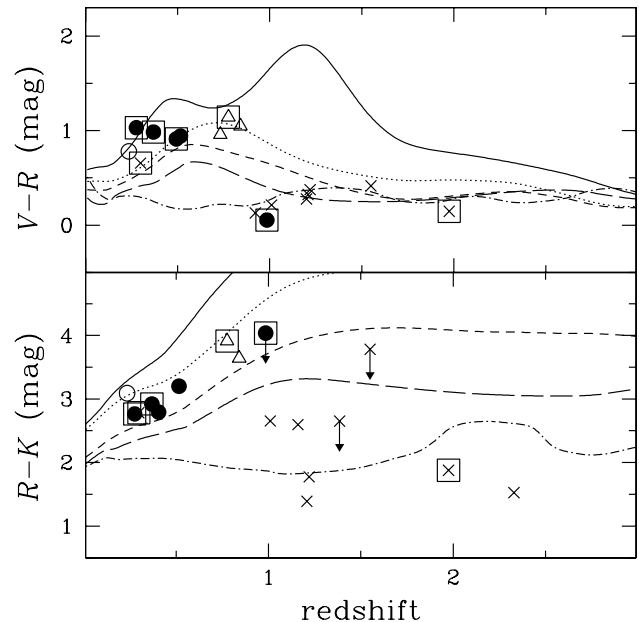,width=3.5in,angle=0}} 
\caption {$V-R$ (top panel) and $R-K$ colour (bottom panel) against 
 redshift. The symbols are the same as in Figure \ref{fig_fxfo}. 
 The curves are different template SEDs for E/S0 (continuous), Sbc
 (dotted), Scd (short dashed), irregulars (long dashed) and QSOs
 (dot-dashed). The galaxy templates are observed SEDs from Coleman, Wu
 \& Weedman (1980). The QSO template is obtained from the set of QSO
 SEDs of  the {\sc le phare} software. The majority of the sources
 with the harder X-ray spectra have optical/NIR  colours consistent with those
 of galaxies.
 }\label{fig_colour_vs_z}    
\end{figure}

We further compare the X-ray and optical/NIR properties of the present
sample in Figure \ref{fig_hr_rk} plotting the  hardness ratio against
$R-K$ colours. Although, there is large scatter in this Figure, the
obscured (high HR) X-ray sources are, on average, redder than those
with softer X-ray spectral properties. This suggests  that the
optical/NIR light of X-ray harder sources is dominated by  the host
galaxy rather than the  obscured central AGN (Barger et al. 2002,
2003; Brusa et al. 2003; Fiore et al. 2003; Mignoli et al. 2004;
Gandhi et  al. 2004; Georgantopoulos et al. 2004). This is
demonstrated in Figure \ref{fig_colour_vs_z} where we plot $R-K$ and
$V-R$ colours as a function of redshift.  Overlaid are the optical/NIR
colours of a QSO spectrum   (Cristiani \& Vio 1990; Cristiani et
al. 2004; obtained from the template SEDs of the 
{\sc le phare}
software\footnote{http://www.lam.oamp.fr/arnouts/LE\_PHARE.html})  and
the mean observed spectra of four different galaxy types (E/S0, Sbc, 
Scd, Im) from Coleman, Wu \& Weedman (1980). Broad line AGNs, most of
which exhibit soft X-ray spectra, have colours consistent with the QSO
template prediction, while the X-ray harder sources follow the galaxy
tracks in Figure \ref{fig_colour_vs_z}. The evidence above justifies
the use of galaxy SEDs to estimate  photometric redshifts for the
X-ray harder sources in section \ref{sample}.     

\begin{figure} 
\centerline{\psfig{figure=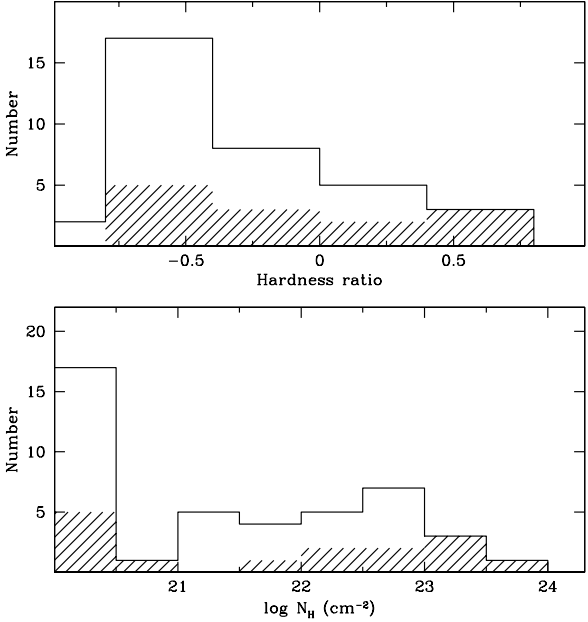,width=3.5in,angle=0}} 
\caption
 {{\bf Top:} Hardness ratio distribution of the hard X-ray selected
 sample (open histogram) in comparison with that of X-ray/radio
 matches (shaded histogram). {\bf Bottom:} rest frame column density
 distribution of the hard X-ray selected
 sample (open histogram) in comparison with that of X-ray/radio
 matches (shaded histogram). Both plots suggest an increasing fraction
 of radio emitting hard X-ray sources with increasing hardness ratios 
 or absorbing column density.  
 }\label{fig_hr_nh_dist}    
\end{figure}

In Figures \ref{fig_hr_vs_fxfo}, \ref{fig_hr_rk} 
there is evidence for a higher fraction of X-ray/radio matches within
the harder (i.e. higher hardness ratio) X-ray population.  This is
further explored in Figure \ref{fig_hr_nh_dist}, where we plot the
distribution of the HR and the {\it rest-frame} column density (i.e.
after correcting for the redshift) of both the hard X-ray selected
sample and the X-ray/radio matched population.  The $\rm N_H$ at the
{\it observer's frame} is estimated from the $\chi^{2}$ X-ray spectral
fittings described in section \ref{sample}. In the case of X-ray
spectra with small number of counts we adopt the $\rm N_H$ values
estimated by the C-statistic method assuming a spectral index
$\Gamma=1.7$. The {\it observer's frame} column density however, is
lower than the {\it rest-frame} one because the $k$-effect shifts the
absorption turnover to lower energies. The relation between the
intrinsic rest-frame and the observed column density scales
approximately as $(1+z)^{2.65}$ (e.g. Barger et al. 2002). This correction
is applied to all the sources in the sample before plotting the
histogram in Figure \ref{fig_hr_nh_dist}. For sources without
spectroscopic identification we assume a mean redshift $z=1$, similar
to the peak of the redshift distribution of the hard X-ray population
(e.g. Fiore et al. 2003). Also, sources with X-ray spectra consistent
with no absorption above the Galactic are plotted at the Galactic
column density, $\rm N_H=2\times10^{20}\,cm^{-2}$.

Figure \ref{fig_hr_nh_dist} suggests that the fraction  of X-ray/radio
matches  increases with the HR or the $\rm N_H$. About 18 per cent
(4/22) of the population with $\rm HR<-0.4$ is associated with radio
emission while 50 per cent (10/20) of the $\rm HR>-0.4$ sources have
radio counterparts.  Similarly, the fraction of X-ray/radio matches is
about 23 per cent (6/26) of the population with  rest-frame column density
$\rm N_H < 10^{22}\, cm^{-2}$ and increases to about 50 per cent
(8/16) for  $\rm  N_H > 10^{22}\, cm^{-2}$. 
However, the small sample size may bias our
conclusions. We therefore compare the HR and $\rm N_H$  distributions 
of X-ray selected AGNs with and without radio counterparts using the 
Gehan's statistical test as implemented in the {\sc asurv} 
package (Isobe, Feigelson \& Nelson 1986; LaValley, Isobe \& Feigelson
1992). The probability the two  distributions are drawn from the same
parent population is rejected at the $\approx95$ per cent confidence
level corresponding to about $2\sigma$.

\begin{figure} 
\vspace{3.5in}
\caption
{The merged X-ray spectrum of the 3 heavily obscured  (rest frame
 $\rm N_H>10^{22}\,cm^{-2}$) radio emitting AGNs at $z\approx0.4$
 (sources \#7, 22 and 42 in Table \ref{tbl1}). For clarity we only
 show the PN coadded spectrum. The continuous line is the single
 absorbed power-law (wabs*pow) model. There is clearly an excess
 above the single component model at energies
 $<2$\,keV.}\label{fig_xspec}     
\end{figure}

\subsection{X-ray spectra}
In this section we focus on the X-ray spectra of both individual
sources and different groups of X-ray selected AGNs.  Firstly, we
compare the X-ray spectral properties of the AGNs with and without
radio counterparts, comprising 14 and 28 systems respectively
The individual spectra of sources in these sub-samples are merged using
the {\sc mathpha} task of {\sc ftools} to produce 3 independent
coadded spectral files for the PN, MOS1 and MOS2 detectors
respectively. The combined spectra are grouped to a minimum of 15
counts per bin to ensure that Gaussian statistics apply. The auxiliary
files of individual sources were combined using the {\sc addarf} task
of {\sc ftools}. Using the {\sc xspec} v11.2 software, we fit a single
power-law to the data absorbed by the Galactic column of $\rm
2\times 10^{20}\,cm^{-2}$ (wabs*pow). The results are presented in Table 
\ref{tbl_res1}. X-ray/radio matched AGNs have flatter spectra 
($\Gamma=1.78^{+0.05}_{-0.03}$) than non-radio detected sources
($\Gamma=2.00^{+0.03}_{-0.04}$) at the $\approx4\sigma$ significance
level. A similar result was obtained by Bauer et al. (2002), who
explored the association between the faint X-ray (0.5-8\,keV) and
radio (1.4\,GHz) source populations detected in the Hubble Deep Field
North region using the 1\,Ms {\it Chandra} dataset and ultra-deep VLA
observations. Although their sample is dominated by starbursts, they
also identify a number of X-ray selected AGNs and argue that those
with radio detections have harder X-ray spectral properties than radio
undetected ones. Bauer et al. (2002) suggest that the enhanced
absorption observed in radio detected AGNs is due to nuclear
starbursts. 

We explore this scenario using X-ray spectral fitting analysis of the
hard (rest-frame $\rm N_H>10^{22}\,cm^{-2}$) X-ray/radio matched
sources in the Phoenix/XMM-{\it Newton} survey.  Although individual
sources have a small number of counts, which do not allow detailed
spectral analysis, we can combine their X-ray spectra to study their
mean properties. The different redshifts of individual
sources may however, dilute the spectral features in the coadded X-ray
spectrum. To avoid this effect we only combine the X-ray spectra of
sources \#7, \#22 and \#42 in Table \ref{tbl1} that lie at similar
redshifts ($\approx0.4$).  The individual PN X-ray spectra of these 
3 sources are shown in Figure \ref{fig_xspec_app} of the Appendix.

The 3 spectra are merged using the method described above. We fit a
single absorbed power-law (wabs*pow) to the data with both $\Gamma$ and
$\rm N_H$ free parameters. We find $\Gamma\approx1.9$, $\rm 
N_H\approx7\times10^{22}\,cm^{-2}$ (not corrected for redshift) and
reduced $\chi^2=2.02$ for 27 degrees of freedom. The results are shown
in Figure \ref{fig_xspec}. This model does not provide a good fit to
data and cannot account for the soft excess below about 2\,keV in
Figure \ref{fig_xspec}. We add a second absorbed power law component
(wabs*pow+wabs*pow) with fixed $\Gamma=1.9$ and absorbing column tied to the
Galactic value $\rm N_H \approx 2\times 10^{20} \,cm^{-2}$. We also
fix the power-law index of the first component to $\Gamma=1.9$,
leaving $\rm N_H$ a free parameter. This gives $\rm
N_H\approx8.9\times10^{22}\,cm^{-2}$ 
(observer's frame) and reduced $\chi^2=1.28$ for 27 degrees of
freedom.  We use the F-test to compare the single and two component
models above and find that the probability of the two models being an
equally good representation of the data is $8.4\times10^{-4}$. The 
single component model can therefore be rejected at 99.92 per cent
confidence level corresponding to about $2.8\sigma$. If we use an
absorbed Raymond-Smith model (wabs*ray+wabs*pow) with a temperature of
0.7\,keV (Franceschini et al. 2003) as the second component, we
estimate a lower F-test probability ($1.6\times10^{-2}$) corresponding
to a rejection significance of $\approx2.4\sigma$. 

The results above are summarised in Table \ref{tbl_res2} and  suggest
the presence of a soft component in at least the low-$z$ sub-sample of
the X-ray/radio  matched AGNs, albeit at the $\approx2.5\sigma$
level. This component may be associated with scattered emission that
is frequently observed in local obscured Seyfert-2 type AGNs
(e.g. Turner et al. 1997). In this picture there is a thick screen of
obscuring material which swamps the soft X-rays while the more
energetic hard X-rays can penetrate through. The soft X-ray emission
arises from scattering on a pure electron medium. 

Alternatively, the soft component may be associated with
star-formation activity. We estimate that this component is
responsible for about 10 per cent of the 0.5-8\,keV luminosity of the
stacked spectrum. Since the 3 low-$z$ X-ray/radio matched sources have
luminosities of $\approx 10^{43}\,\rm  erg\,s^{-1}$ , the above
fraction translates to a 0.5-8\,keV luminosity of about $10^{42}\,\rm
erg\,s^{-1}$ consistent with that of extreme starbursts in the local
Universe (e.g. Moran, Lehnert \& Helfand 1999;  Georgakakis et
al. 2003). We also estimate an upper limit to the star-formation rate
of $\rm SFR=100\,M_{\odot}\,yr^{-1}$ for these systems,  using the
relation between radio luminosity density and SFR (Bell et
a. 2003). Any ongoing  SFR is most likely lower than the limit above
since the AGN will also contribute to the observed radio emission. A
SFR of $\rm 100\,M_{\odot}\,yr^{-1}$ is nevertheless, not unreasonably
high for a starburst galaxy (e.g. Moran et al. 1999; Ranalli, Comastri
\& Setti  2003).  

\begin{table} 
\footnotesize
\begin{center} 
\begin{tabular}{cc cc c} 
\hline                        

AGN     & number of & $\Gamma$ & $\chi^2$ & d.o.f. \\
sample  &  sources  &          &         &        \\
\hline
X-ray/radio  & 14  & $1.78^{+0.05}_{-0.03}$  & 1.17 & 314 \\
X-ray   & 28  & $2.00^{+0.03}_{-0.04}$  & 1.15 & 431 \\
\hline
\end{tabular}
\end{center}
\caption{
The mean X-ray spectral properties of the X-ray selected AGNs with 
and without radio identifications. The X-ray spectra of the two
subsamples are fit by an absorbed power-law with the $\rm N_H$ fixed
to the Galactic value. 
}\label{tbl_res1}
\normalsize
\end{table}

\begin{table*} 
\footnotesize
\begin{center} 
\begin{tabular}{cc cc cc} 
\hline                        
\multicolumn{2}{c}{1st component}   &  \multicolumn{2}{c}{2nd component}  & $\chi^2$ & d.o.f. \\
\hline

\multicolumn{2}{c}{power-law}   &  \multicolumn{2}{c}{---}    &   &   \\
$\Gamma=1.86^{+0.41}_{-0.44}$    & $\rm N_H=7.0^{+1.8}_{-1.7}\times10^{22}\, cm^{-2}$  & \multicolumn{2}{c}{---}  & 2.02 & 27 \\ 
\multicolumn{2}{c}{power-law}   &  \multicolumn{2}{c}{power law}    &   &   \\

$\Gamma=1.9$ (fixed) & $\rm N_H=8.9^{+2.7}_{-2.0}\times10^{22}\, cm^{-2}$
& $\Gamma=1.9$ (fixed) & $\rm N_H=2\times10^{20}\, cm^{-2}$ (fixed)  & 1.28
& 27 \\  

\multicolumn{2}{c}{power-law}   &
\multicolumn{2}{c}{Raymond-Smith}    &   &   \\ 

$\Gamma=1.9$ (fixed) & $\rm N_H=7.4^{+2.2}_{-1.5}\times10^{22}\, cm^{-2}$  & kT=0.7
(fixed) & $\rm N_H=2\times10^{20}\, cm^{-2}$ (fixed)  & 1.417 & 26 \\  
\hline
\end{tabular}
\end{center}
\caption{
The mean X-ray spectral properties of the 3 X-ray sources with radio
counterparts that lie at similar redshifts ($z\approx0.4$). Different
model components are used to fit the coadded spectrum.  
}\label{tbl_res2}
\normalsize
\end{table*}

\subsection{Contribution to the XRB}
Summing the X-ray fluxes of individual sources (after correcting for
vignetting) from $f_X (\rm 2 - 8 \, keV ) = 1.24\times10^{-13}\rm \,
erg \, s^{-1} \, cm^{-2}$ (i.e. brightest source) to the limit $f_X
(\rm 2 - 8 \, keV ) = 7.7 \times 10^{-15}\rm \, erg \, s^{-1} \,
cm^{-2}$, we estimate a resolved flux of $(6.8\pm1.3)\times
10^{-12}\rm \, erg \, s^{-1}\,deg^{-2}$ assuming Poisson statistics
for the errors.  This corresponds to $38\pm7 - 52\pm10$ per cent of
the   2-8\,keV XRB measured by BeppoSAX (Vecchi et al. 1999) and ASCA
(Ueda et al. 1999) respectively.  For sources brighter than $f_X (\rm
2 - 8 \, keV ) = 1.24\times10^{-13}\rm \, erg \, s^{-1} \, cm^{-2}$
(i.e. not probed in our relatively small solid angle survey) we
extrapolate the Baldi et al. (2002) best fit relation to estimate a
contribution of $\approx3\times10^{-12}\rm \, erg \, s^{-1}\,deg^{-2}$
or about 16--23 per cent of the total XRB.

For the X-ray sources with radio counterparts we estimate a total flux
contribution from the  X-ray/radio matched population of
$(2.6\pm11.0)\times 10^{-12}\rm \, erg \, s^{-1}\,deg^{-2}$ or $14\pm6
- 20\pm8$ per cent of the XRB for the BeppoSAX and ASCA normalisations
respectively. X-ray/radio matched sources represent about $38\pm16$
per cent of the XRB fraction resolved in our survey for fluxes in the
range $f_X (\rm 2 - 8 \, keV ) = 7.7  \times 10^{-15} - 1.24 \times
10^{-13} rm \,   erg \, s^{-1} \, cm^{-2}$ of the present
survey. Radio emitting AGNs are therefore, a non-negligible fraction
of the X-ray background. We also note, that this fraction is skewed
by the bright sources in the sample. Excluding the two brightest
sources (both of which show radio emission) we estimate that the 
X-ray/radio matched population sums up to about $24\pm8$ per cent 
of the XRB resolved in the flux range of our survey. Within the
$1\sigma$ uncertainties however, this is in agreement with the
fraction estimated above taking into account the bright sources in the
sample.

Barger et al. (2001) combined a Chandra survey of the SSA13 region
(limiting flux $f_X(\rm 2 - 10 \, keV)= 4 \times 10^{-15} \, erg \,
s^{-1} \, cm^{-2}$) with deep radio observations to a $5\sigma$
limiting flux density of $\rm 25\,\mu Jy$. They estimate a
contribution of the X-ray/radio matched population to the XRB of $\rm
\approx 26$ per cent. Although this is somewhat higher than the
fraction estimated above (14-20 per cent), the difference is likely
due to their deeper radio and X-ray data when compared to the 
Phoenix/XMM-{\it Newton} survey.

\section{Discussion}\label{discussion}

Using the Phoenix/XMM-{\it Newton} survey, we have presented evidence
for a higher fraction of X-ray/radio matches among the obscured
AGN population compared to sources with softer X-ray spectral
properties. Small number statistics however, limit the significance of
this result to the $2\sigma$ level.  

Previous studies also suggest a higher fraction of
obscured AGNs within the X-ray/radio matched population.  Ciliegi et
al. (2003) find that the fraction of X-ray sources with radio
counterparts is higher in hard rather than soft X-ray selected
samples. This is attributed to both observational effects (e.g. deeper
radio data compared to the X-ray flux limit for the harder samples)
and the hard band selection.  They argue that both the hard X-ray
energies and the radio wavelengths are least biased by dust. Selection 
at radio and hard X-ray wavelengths therefore provides more complete
AGN samples that include obscured AGNs likely to be missed in soft
X-ray surveys. Similarly, Barger et al. (2001) found that as much as
50 per cent of their hard X-ray sources to the limit $f_X(\rm 2 - 10
\, keV)= 4 \times 10^{-15} \, erg \, s^{-1} \, cm^{-2}$ have radio
counterparts above the $5\sigma$ flux density level of $S_{1.4}=\rm 25
\, \mu Jy$. Moreover about half of their X-ray/radio matched sources
show flat X-ray spectra ($\Gamma\la1$) suggesting enhanced
photoelectric absorption. 

A circum-nuclear starburst is  an alternative scenario for the higher 
fraction of X-ray/radio matches within the X-ray harder
population. Indeed, since the sub-mJy radio data are sensitive 
to star-formation activity it may be possible that the hard
X-ray/radio matches are composite systems comprising both a central
AGN and a nuclear starburst that both feeds and obscures the central
engine (e.g. Fabian et al. 1998).   

The coadded X-ray spectra of the 3 low-$z$ hard X-ray selected sources
with radio counterparts show evidence for the presence of a soft
component, albeit at the $\approx2.5\sigma$  level. Although this soft
excess may be associated with reflection on the dusty torus (Reeves et
al. 1997) or scattering on a pure electron medium (Turner et al. 1997)
an alternative scenario is star-formation activity.  In the local
Universe there is a wealth of observational data that suggest a link
between circum-nuclear starbursts and  AGNs. For example the
featureless UV continua of Seyfert 2s have been proposed to be largely
produced by a nuclear starburst (Heckman et al. 1995, 1997;
Gonzalez-Delgado et al. 1998). More recently, Gonzalez Delgado et
al. (2001) and Storchi-Bergmann et al. (2000) estimated that about
30-50 per cent of their Seyferts 2 samples have near-UV spectra that
show evidence for the existence  of nuclear starbursts. Similarly,
the identification of the Ca\,II triplet in  the NIR  spectra of
Seyfert 2s  has been interpreted as evidence for the presence of young
stars that are likely to dominate the NIR emission of these systems 
(e.g. Terlevich, Diaz \& Terlevich 1990). 

The radio and FIR properties of many Seyfert galaxies are  also
consistent  with starburst activity (Norris, Allen \& Roche
1988). Bransford et al. (1998) showed that about 40 per cent of their
sample,  comprising mostly Seyfert 2s, has radio morphology and/or
radio-to-FIR flux ratios suggesting star-formation in addition to the
central AGN activity. Ulverstad \& Ho (2001) explored the FIR and
radio properties of Seyfert 1 and 2s compiled from the Palomar
optical spectroscopic survey of nearby galaxies (Ho, Filippenko \&
Sargent 1995). They find that the radio-to-FIR flux ratio of many
Seyfert 2s in their sample, particularly the less luminous ones, is
consistent with star-formation. This is somewhat contradictory to the  
conclusions of Cid Fernades et al. (2001) and Kauffman et al. (2003) 
who argue that it is the most powerful (e.g. more far-infrared
luminous) Seyfert 2s that show evidence for the presence of young
stars.   

Moreover, although a number of nearby galaxies (e.g. NGC\,6240,
NGC\,4945) have properties consistent with starburst/LINER activity,
observations at hard  X-ray wavelengths have revealed a heavily
obscured AGN that is hidden from view at any other wavelength (Iwasawa
et al. 1993; Iwasawa \& Comastri 1998). Such  ``composite''
(starburst+AGN) objects  have also been proposed as a significant mode
of the XRB (Fabian et al. 1998).    
    
Studies of the X-ray properties of luminous infrared galaxies (LIGs;
$L_{FIR}>10^{11}\, L_{\odot}$), believed to be the local counterparts
of the more distant sub-mJy and $\mu$Jy radio population, also suggest
a close link between star-formation and AGN activity. Risaliti et
al. (2000) propose a model that invokes a mixture of obscured AGN and
starburst activity to reproduce the  FIR and X-ray  properties of 
LIGs. Franceschini et al. (2003) used the XMM-{\it Newton} data to 
investigate the X-ray spectral properties of a small sample of 10
ultra-luminous infrared galaxies (ULIGs; $L_{FIR}>10^{12}\,
L_{\odot}$). All the systems in their sample show evidence for thermal
hot plasma emission below $\approx1$\,keV likely to be associated with
a nuclear or a circum-nuclear starburst in agreement with our
results. They also find that about half of the sources have X-ray
spectral properties suggesting heavily obscured AGN activity.  

At higher redshifts Bauer et al. (2002) also suggested the presence of
nuclear star-formation activity in X-ray selected AGNs with radio
counterparts on the basis of both their X-ray and radio properties.
Additionally they show that X-ray/radio matched AGNs have flatter X-ray
spectra than AGNs without radio emission in agreement with our
results.  Radio loud AGNs have also been found to have flatter X-ray
spectral properties ($\Gamma=1.6$) than radio quiet ones (Reeves \&
Turner 2000; Gambill et al. 2003). We note, however, that most of the
X-ray/radio matched sources in the present study are either
radio-quiet or lie on the borderline between radio-loud and
radio-quiet systems. Furthermore, the majority of the sources in the
studies above are more luminous at both X-ray, optical and radio
wavelengths than the systems presented here. Most show broad optical
emission lines, while our X-ray/radio matched population with hard
X-ray spectral properties and available optical spectroscopy have
narrow optical emission lines. This suggests that the absorbed
population studied here is different from the radio loud QSO samples
of Reeves \& Turner (2000) and Gambill et al.  (2003).

\section{Conclusions}\label{conclusions}
In this paper we have explored the radio properties of hard X-ray
selected sources using a deep (50\,ks) XMM-{\it Newton} pointing
overlapping with a subregion of a deep and homogeneous radio survey
(the Phoenix Deep Survey) reaching $\mu$Jy sensitivities. A total of
43 sources are detected above the X-ray flux limit $f_X(\rm 2 - 8 \,
keV) = 7.7 \times 10^{-15}\, erg\, s^{-1}\, cm^{-2}$, with 14 associated
with radio sources. A total of 29 hard X-ray selected sources are
optically identified using either relatively shallow ($R=22.5$\,mag)
or deeper ($R\approx24$\,mag) optical data.  Spectroscopic data are
available for 18 of the optically brighter X-ray sources.

The hard X-ray selected sample comprises  (i) 11 AGNs with broad
emission lines, 2 of which are associated with a radio source (ii) 6
systems with narrow emission line optical spectra, 4 of which also have
radio emission, (iii) 1 galaxy with absorption optical lines, (iv) 1
Galactic star candidate on the basis of its optical and X-ray
properties, (v) 12 sources with optical identification but no
spectroscopic redshift measurement and (vi) 12 optically unidentified
sources, 2 of which are associated with a radio source. One of the
narrow  emission line X-ray/radio sources may be a candidate type-2
QSO at $z\approx0.985$.  

We find evidence for an increasing fraction of X-ray/radio matches
with increasing hardness ratio or rest-frame column density,
suggesting that radio detected AGNs have, on average, harder X-ray
spectral properties. Indeed, X-ray selected AGNs associated with radio
counterparts have flatter X-ray spectra than the non radio detected
X-ray sources. We argue that the enhanced photoelectric absorption
observed in radio emitting X-ray selected AGNs  is likely to be
associated with circum-nuclear starburst activity that both feeds and
obscures the central engine. At least for a small sub-sample of  
low-$z$ radio emitting AGNs their combined spectrum exhibits a soft
X-ray component, albeit at the $2.5\sigma$ confidence level,  that may
be associated with star-formation activity. However, we cannot exclude
the reflection on the dusty torus or the scattering on a pure electron  
medium as alternative scenarios for the observed soft emission. 
The unparalleled
sensitivity of the {\it Spitzer} observatory can potentially provide
more information on the presence of a circum-nuclear starburst in
these radio emitting AGNs. Mid-infrared spectroscopy can provide useful
diagnostics on the nature of the central source (e.g. Genzel \&
Cesarsky 2000) while, mid- to far-infrared photometry can  constrain
the SEDs of these systems allowing one to assess the relative
contribution of the starburst and the AGN components (e.g. Farrah et
al. 2003). Deeper X-ray observations are also essential to improve the
photon statistics and to perform detailed X-ray spectral analysis of
these sources.   

The optical and NIR colours of the harder X-ray sources are consistent
with those of galaxies, suggesting that the optical and the NIR light
of this population is dominated by the host galaxy rather than the
obscured AGN. By contrast, X-ray sources with soft X-ray spectral
properties (many of which are identified with broad optical emission
lines) have blue colours consistent with those of Seyfert 1s and QSOs.

Finally, to the limit of the present radio and X-ray data,
we find radio emitting AGNs make up a non-negligible fraction of
the XRB. About $14\pm6 - 20\pm8$ per cent of the total XRB for the
BeppoSAX and ASCA normalisations 
respectively arises in AGNs associated with radio emission. 
About half of this fraction arises from sources with hard X-ray
spectral properties, suggesting enhanced absorption.  

\section{Acknowledgments}
The authors wish to thank the anonymous referee for constructive
comments, Ioannis Georgantopoulos for numerous discussions and for
his useful suggestions and Athanassios Akylas for his help with the
X-ray spectral fittings.  The Phoenix Deep Survey radio data as well
as part of the observations presented here are electronically
available at {\sf http://www.atnf.csiro.au/people/ahopkins/phoenix/}.
AG acknowledges funding by the European Union and the Greek Ministry
of Development in the framework of the Programme 'Competitiveness --
Promotion of Excellence in Technological Development and Research--
Action 3.3.1', Project 'X-ray Astrophysics with ESA's mission XMM',
MIS-64564. AMH acknowledges support provided by the National
Aeronautics and Space Administration through Hubble Fellowship grant
HST-HF-01140.01-A awarded by the Space Telescope Science Institute. JA
gratefully acknowledges the support from the Science and Technology
Foundation (FCT, Portugal) through the fellowship BPD-5535-2001 and
the research grant POCTI-FNU-43805-2001. MS acknowledges support from
a PPARC fellowship.

\appendix
\section{Notes on selected sources}\label{app1}
In this section we discuss in more detail the properties of the
X-ray sources with radio identification and available spectroscopic 
information. In addition to these sources we also discuss the
properties of the candidate star (source \#25) identified in our
sample.   

{\bf \#7:} This source has a narrow emission line optical
spectrum at  $z=0.276$ (see Figure \ref{fig_spec}).  The relative
strengths  of the $\rm H\beta$ and $\rm [O\,III]\,5007\AA$ emission
lines as well as the presence of high excitation lines (e.g $\rm
[Ne\,III]\,3869$, $\rm [Ne\,V]\,3426\,\AA$) suggest obscured AGN
activity. This is  consistent with the flat X-ray spectrum
suggesting a rest frame absorbing column of $\rm
\approx 4\times10^{22}\,cm^{-2}$ ($\Gamma=1.7$). The optical $R$-band
image with the X-ray and the  radio positions overlaid are shown in
Figure \ref{fig_image}. Visual inspection of the $R$-band image shows
two components likely to be in the process  of interaction/merging.

{\bf \#12:} This source is assigned a redshift $z=0.985$ on the basis
of a single narrow emission line interpreted as $\rm
[O\,II]\,3727\,\AA$.  The X-ray spectrum suggests a high column
density  $\rm N_H=1.3\times 10^{23} \,cm^{-2}$
($\Gamma=1.7$). The X-ray luminosity of this source at $z=0.985$
before applying any correction due to intrinsic photoelectric
absorption is estimated  $L_X(\rm 2 - 8 \,keV)\approx10^{44} \, erg \,
s^{-1}$. The evidence above suggests a candidate type-2
QSO. Interestingly, the optical colour of this source is very blue
$V-R\approx0$\,mag. This is hard to reconcile with the heavily
absorbed X-ray spectrum suggesting significant amounts of dust. It is
possible that the very blue optical colour is associated with
starburst activity in the host galaxy. The optical spectrum and the
optical $R$-band image with the X-ray and the radio positions overlaid
are shown in  Figures  \ref{fig_spec} and  \ref{fig_image}
respectively.     

{\bf \#13,  23:} These 2 sources have broad emission line systems
with soft X-ray spectral properties. They are associated with radio
sources, albeit below the formal $\rm 80\mu Jy$ radio detection
level corresponding to about $5\sigma$. The optical spectra and the
optical images of these sources are shown in Figures \ref{fig_spec}
and \ref{fig_image} respectively.

{\bf \#22:} This source has properties very similar to those of
\#7. At $z=0.369$ it also has a narrow emission line optical spectrum
(see Figure \ref{fig_spec}) with relative line strengths suggesting
obscured AGN  activity. This is also consistent with the hard X-ray
spectral properties of this source suggesting a rest-frame column
density $\rm N_H=1.6\times 10^{23} \,cm^{-2}$. The
optical $R$-band image of this source with X-ray and the radio
positions overlaid are also shown in  Figure \ref{fig_image}.

{\bf \#25:} This X-ray source is associated with an unresolved bright
$R\approx13$\,mag optical source, most likely a Galactic star. The
observed X-ray--to--optical flux ratio is estimated $\approx -3$, while
the X-ray spectrum of this source is best fit by a Raymond-Smith model
with $\rm kT=0.70^{+0.06}_{-0.05}$ and abundance
$0.038^{+0.017}_{-0.010}$ ($\chi^{2}=1.5$ for 81 degrees of
freedom). This is source is not shown in  Figure \ref{fig_image}
although its X-ray spectrum is plotted in \ref{fig_xspec_app}

{\bf \#32:} This X-ray source is associated with a pair of optical
galaxies likely to be interacting (see Figure
\ref{fig_image}). This source has been  observed by HYDRA yielding the 
featureless spectrum shown in Figure \ref{fig_spec} that did not allow
a redshift determination. The evidence above in combination with the soft
X-ray spectral properties of source \#32 may suggest a
BL-Lac. However, the two point radio/optical and X-ray/optical
indices of this source ($\alpha_{OX}=1.02$, $\alpha_{RO}=0.09$) are not
consistent with those expected for BL-Lacs (Stocke et al. 1991). 

{\bf \#42:} This source has a narrow emission line spectrum presented
in Figure \ref{fig_spec}. The relative strengths  of the $\rm H\beta$
and $\rm [O\,III]\,5007\AA$ emission lines suggest obscured AGN 
activity. This is also confirmed by the hard X-ray spectral
properties of this source suggesting an rest-frame column density of
$\rm N_H\approx2\times 10^{23} \,cm^{-2}$ ($\Gamma=1.7$). The optical
$R$-band image of this source with X-ray and the radio positions
overlaid are shown in  Figure \ref{fig_image}.

\begin{figure*} 
\vspace{7in}
\caption
 {Optical spectra of the  X-ray/radio matches. The intensity shown is
 the observed count rate, since we have not performed flux calibration
 (see section 
 \ref{sec_spec}).  
 }\label{fig_spec}
\end{figure*}

\begin{figure*} 
\vspace{7in}
\caption
 {Optical $R$-band images of the X-ray/radio matches. The circle has 
 a radius of 5\,arcsec and is centered at the position of the X-ray
 source centroid. The square has a side of 7\,arcsec and shows the
 radio source position.   
 }\label{fig_image}    
\end{figure*}

\begin{figure*} 
\vspace{6in}
\caption
 {X-ray spectra of individual X-ray/radio matches with optical
 spectroscopy. For clarity we show only the PN spectrum. The curves
 are the wabs*pow models described in section 3 with the parameters
 listed in Table 2. In the case of C-statistic (sources 
 \#7, 12, 22, 42) we fix the exponent $\Gamma$ to 1.7 and fit the
 $N_H$. In the case of $\chi^2$ analysis (sources \#13, 23, 25, 32)
 both the  $\Gamma$ and the  $N_H$ of the wabs*pow model are free
 parameters. The only exception is the candidate Galactic star, source
 \#25, for which we show the best fit Raymond-Smith model. The
 reduced $\chi^2$ for sources \#13, 23, 25, 32 is respectively 1.55,
 0.76, 1.5 and 1.14 for 28, 27, 81 and 224   degrees of freedom
 respectively.  
 }\label{fig_xspec_app}
\end{figure*}

\end{document}